# Effect of multiplicity of stellar encounters and the diffusion coefficients in the uniform stellar medium: no classical divergence ?

## A.S. Rastorguev [1, 2], O.V. Chumak [1], N.D. Utkin [2]

[1] *Lomonosov Moscow State University, Sternberg Astronomical Institute*

[2] *Lomonosov Moscow State University, Faculty of Physics*

**ABSTRACT**

*Agekyan's λ-factor* that accounts for the effect of multiple stellar encounters with large impact factors is used for the first time to compute the diffusion coefficients in the velocity space of a stellar system. It is shown that in this case the cumulative effect – the total contribution of distant encounters to the change in the velocity of the test star – is finite, and the logarithmic divergence inherent to the classical description disappears, as also was earlier noted by Kandrup (1981). The formulas for the diffusion coefficients, as before, contain the logarithm of the ratio of two independent scale factors that fully characterize the state of the stellar system: the average interparticle distance and the impact parameter of a close encounter. However, the physical meaning of this factor is no longer associated with the classical logarithmic divergence.

## Introduction

The current description of collisional processes in the dynamics of stellar systems dates back to the pioneering studies of the first half of the 20th century (Charlier 1917, Jeans 1919, Spitzer 1924, Roseland 1928, Smart 1938, Williamson and Chandrasekhar 1941, and Chandrasekhar 1941a, b, 1942). An exhaustive historical review of the works on the kinetic theory of uniform systems with long-range interactions including stellar systems can be found in the extensive study by Chavanis (2013). Chandrasekhar (1941a) was the first to apply Holtsmark's (1919) distribution of random force to uniform stellar media, and show that the asymptotics of this distribution in the limit of large forces coincides with the distribution of the force due to the nearest neighbor (Hertz, 1909). This very important property of gravitating systems forms the basis of the approach that allows collisional kinetics of stellar medium, i.e., the variation of stellar velocities, to be described in terms of the concept of two-particle encounters. In fact, significant random changes in the velocity of a test star occur only during sufficiently close encounters between the test star and field stars at characteristic distances that are significantly smaller than the average interparticle distance. The combined effect of such successive encounters and the corresponding diffusion coefficients in the velocity space can then be computed, e.g., in terms of the approach based on the concept of independent random process (as a cumulative effect of stellar encounters). The subsequent development of stellar dynamics in the second half of the 20th century was based entirely on these basic principles.

There are several methods for estimating the cumulative effect of stellar encounters: by the deflections of the velocity vector of the test star (Williamson and Chandrasekhar, 1941; Parenago, 1954), by the change of the parallel (dynamic friction) and normal (diffusion, scattering) components of the star's velocity (Chandrasekhar, 1941a; King, 2002; Binney and Tremaine, 2008, etc.). All estimates of the rate of change of the velocity and kinetic energy usually yield quite similar values for the time scale of kinetic processes, which is usually identified with the collisional relaxation time. A characteristic feature of the diffusion-coefficient and relaxation-time estimates is the logarithmic divergence, which arises at the upper integration limit of the cumulative effect of two-particle encounters over impact parameter. It has the form of the term $\Lambda = \ln \frac{d_{max}}{d_{90}}$ – an analog of the so-called Coulomb logarithm in plasma physics – that appears in the formulas for the diffusion coefficient. Here $d_{max}$ is the upper limit of impact parameter; $d_{90} = \frac{G \cdot (m+m_f)}{V_0^2}$, the impact parameter of a close encounter, where the vector of relative velocity of two approaching stars is deflected by 90°; $G$, the gravitational constant; $m$ and $m_f$, the masses of the test and field star respectively, and $V_0$, the magnitude of the relative velocity vector.

The problem of the upper limit of impact parameter has been raised repeatedly by many authors in their studies of stellar dynamics. Thus Williamson and Chandrasekhar (1941) and Parenago (1954) pointed out that in the concept of two-particle encounters the natural upper limit for the impact parameter should be equal to the average interparticle distance $\bar{d} \approx 0.554 \cdot \nu^{-1/3}$ (here $\nu$ is the stellar number density), because all the weaker encounters are actually multiple, implying that integration overestimates their combined effect. We share this viewpoint and note that by treating such encounters as involving only two particles we actually incorporate into the resulting diffusion coefficients not only the effect of irregular forces, but also, to a certain extent, the effect due to the regular component of the gravitational field. Ambartsumyan (1938), Ogorodnikov (1958), Kandrup (1980, 1981), Binney and Tremaine (2008), and other authors mentioned in the latter monograph, on the contrary, believe that $d_{max}$ should be set equal to the characteristic size of the entire stellar system (the radius of the cluster, thickness of the galactic disk) or the radius of the regular stellar orbit. Note that the precise knowledge of the upper integration limit is by no means critical for practical purposes (estimation of the relaxation time and computation of the diffusion coefficients), because the rather weak (logarithmic) divergence cannot change significantly the estimates of the above quantities whatever a realistic choice of the maximum impact parameter. Indeed, we have for the solar neighborhood in the Galaxy $\bar{d} \approx 1\ pc$, $d_{90} \approx 1 - 2\ AU$, and $\Lambda \approx \ln \frac{\bar{d}}{d_{90}} \sim 11 - 12$. Adopting $d_{max} \sim H_z \approx 100$ pc as the upper limit increases the "Coulomb logarithm" $\Lambda$ to $\Lambda \sim 15 - 16$, i.e., only by 40-50%, with no radical effect whatsoever on our estimates. However, the problem of choosing the upper limit for impact parameter has another aspect, which is directly associated with the physical basis of collisional kinetics of stellar systems, and we believe that a more in-depth understanding of the physics of such phenomena and attempts to describe them in a noncontroversial way is a task of fundamental importance. These are the issues addressed in this study.

## 1. Multiplicity of stellar encounters

Agekyan (1959) developed and implemented a probabilistic approach to account for stellar encounters, and derived analytical formulas for the probability $\Phi(V^2, h)$ of a stellar encounter producing the given change in the velocity of the test star in some special cases. Here $h = \frac{\Delta V^2}{V^2}$, $\Delta V^2$ is the change in the squared velocity of the star. The weak point of Agekyan's approach is the divergence of the probability for small changes of velocity, $\Phi(V^2, h) \sim h^{-3}$, which, in particular, prevented the computation of the average change in the star's energy. It is evident that this divergence is directly associated with the multiplicity of distant encounters mentioned above, which results in small velocity changes in the computation of the cumulative effect. To attenuate the divergence, Agekyan (1961) introduced a factor accounting for the multiplicity of encounters (Agekyan's *λ-factor*). This factor is equal to the ratio of the magnitude of the random force $|\delta \vec{F}|$ acting on the test star and produced by all stars within a thin spherical layer to the arithmetic sum of the magnitudes of the forces $\sum |\vec{F}_i|$ acting on the test star and produced by all these stars, i.e., $\lambda(p) = \frac{|\delta \vec{F}|}{\sum |\vec{F}_i|} < 1$.

This problem was later discussed by Kandrup (1980), who used simplified approach and suggested that the forces from distant stars effectively cancel each other. Detailed quantitative description of an irregular force field in locally homogeneous stellar field was presented in very important paper of Kandrup (1981). He was the first to note that the diffusion coefficients in the Fokker-Planck approximation do not diverge on the upper limit of integration over the impact parameter.

*Agekyan's λ-factor* has a simple physical meaning. In fact, the actual change in the velocity of the test star (within unit time interval) due to stellar encounters with impact parameters in the *(p, p+dp)* interval is determined by the magnitude of random force, $|\delta \vec{F}|$, i.e., by the geometric sum of the forces produced by all stars in the spherical layer. The use of the arithmetic sum of forces arising in two-particle encounters instead of the random force, as is the case in the calculations of the cumulative effect, results in substantially overestimated values of both the irregular force and the effect of encounters. The need to take into account the attenuation of the effect of distant encounters becomes absolutely clear if one recalls Newton's theorem about a spherically symmetric distribution of external masses. Uniform discrete distribution of gravitating material points evidently has similar properties at large distances from the test particle because particles are distributed practically uniformly over all angles. Scattering centers – i.e., field stars – are distributed randomly and uniformly with average number density *n*, and therefore the vectors of their forces acting onto the test star cancel out. This results in the effect of sui generis total *levelling* of the random force of two-particle encounters already at several interparticle distances. *Agekyan's factor* allows us to compensate the overestimation of the effect of distant encounters. To compute *λ(p)*, Agekyan (1961) used the technique earlier employed to derive Holtsmark's (1919) distribution, and obtained the following formula

$$\lambda(p) = \frac{4}{\pi} \int_0^\infty \frac{x - \sin x}{x^3} \exp\left(-a \frac{4\pi}{3} \nu p^3 x^{3/2}\right) dx, \qquad (1)$$

where $a = \frac{2}{5}\sqrt{2\pi} \approx 1.00265$

Taking into account that $\frac{4\pi}{3} \nu p^3 \equiv N(p)$ – average number of stars inside the sphere of radius $p$, where $p$ is the impact factor of the encounter under consideration – we can rewrite (1) in the equivalent form, treating *λ-factor* as a function of $N = N(p)$:

$$\lambda(N) = \int_0^\infty \frac{x - \sin x}{x^3} exp\left(-a N x^{3/2}\right) dx$$

Thus *Agekyan's λ-factor* is fully determined by the average number of stars located inside the sphere of radius equal to the impact parameter of the encounter considered. We will use this form of the *λ-factor* throughout the text. Function *λ(p)* cannot be expressed in terms of elementary functions, however, at large N is has the well-known asymptotic behavior $\sim N^{-2/3} \sim p^{-2}$ and rapidly decreases with increasing impact parameter. Figure 1 shows the dependence of *Agekyan's λ-factor* on impact parameter expressed in the units of average interparticle distance, $p' = p/\bar{d}$. As is evident from the figure, the effect of stellar encounters is overestimated by one order of magnitude even at two average interparticle distances from the test star. Thus Agekyan quantitatively confirmed the intuitive conclusion of Williamson and Chandrasekhar (1941) that within the framework of 3D Poisson model of the medium the immediate neighborhood of the test particle is the main contributor to the random force.

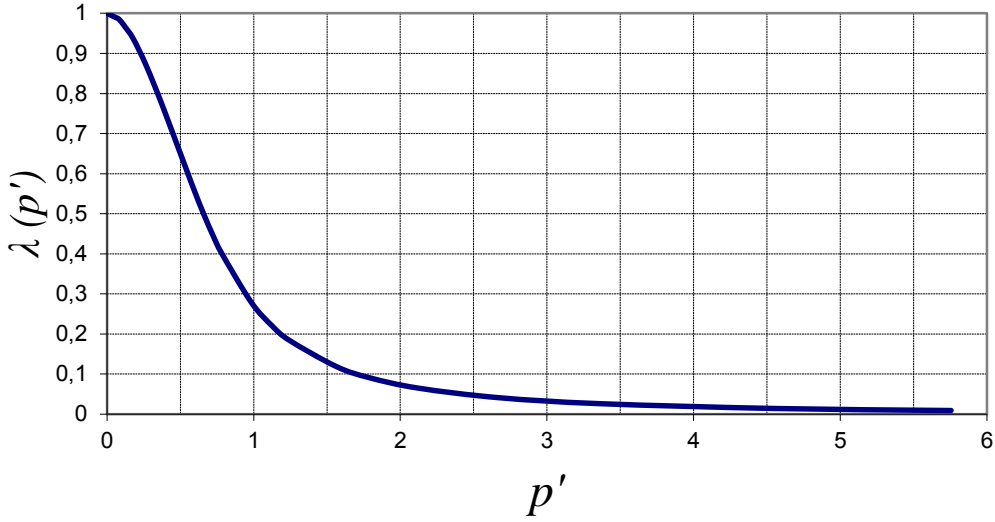

Fig.1. Dependence of *Agekyan's λ-factor* on impact parameter $p'$ expressed in the units of average interparticle distance $\bar{d}$.

Strictly speaking, it would be logical to use *Agekyan's λ-factor* to estimate the contribution of encounters by integrating over impact parameters as it is done within the framework of both the Fokker—Planck approximation and the concept of Markov process described by the Kolmogorov—Feller equation (see below). However, Agekyan (1961) derived an explicit formula for his *λ-factor* several years after he published his fundamental paper where he computed the probability of an encounter with the given velocity change (Agekyan, 1959). It is evident from the technique of derivation described in the paper of Agekyan (1959) that it would be impossible to obtain a finite analytical formula for probability with the allowance for *λ-factor* even for the simplest cases. Realizing this, Agekyan (1961) adopted a palliative decision and introduced a correction factor to the earlier derived formula for the probability of an encounter. The probability corrected to account for the multiplicity of distant encounters has the form $\widetilde{\Phi}(V^2, h) = \lambda(\bar{p}) \cdot \Phi(V^2, h)$, where $\bar{p}$ is the characteristic impact parameter such that the corresponding encounters change the squared velocity by $h$.

Probabilistic approach proved to be highly fruitful, albeit, in our opinion, not enough popular because of the extreme complexity of computations. Thus Petrovskaya (1969 a, b) used it to describe the change in the velocity of a test star in the irregular force field as a purely discontinuous random process. Indeed, introduction of the probability of a stellar encounter with the given change in velocity allows the balance equation for the collisional term to be written in the form of Kolmogorov—Feller integro-differential equation for phase-space density. In their series of papers Kaliberda and Petrovskaya (1970, 1971, 1972) and Kaliberda (1971, 1972) used numerical methods to derive the equilibrium solutions for the Kolmogorov—Feller equation for the local distribution of velocities of stars of different masses. The undeniable advantage of describing collisional kinetics as a purely discontinuous random process compared to the concept of classical diffusion in the velocity space is that the former makes it possible to directly compute not only the dissipation rate, but also the energy losses, which evidently accelerate the dynamical evolution of a star cluster and reduce its lifetime.

## 2. Rigorous account of the multiplicity of encounters in terms of the diffusion approximation

We consider it appropriate and timely to use the fundamental results of Agekyan (1961) to compute directly the diffusion coefficients in a uniform stellar medium with the allowance for the contribution of distant encounters. The task of rigorous account for the multiplicity of stellar encounters in the diffusion approximation does not seem so hopeless as when addressed in terms of the probabilistic approach mentioned. In contrast to Kandrup (1981) approach, we will perform direct integration of the diffusion coefficients over the impact parameters to check if there is no logarithmic divergence.

We take as a basis the derivation of the diffusion coefficients described in the monograph by Binney and Tremaine (2008, Fig. L.6). The initial formulas for the components of the diffusion tensor for a test star averaged over only the orientation angle of the relative orbit have the form:

$$\langle \Delta V_i \rangle = -\Delta V_\parallel \left( \vec{e_i} \cdot \vec{e_1'} \right), \tag{2}$$

$$\langle \Delta V_i \cdot \Delta V_j \rangle = (\Delta V_\parallel)^2 \left( \vec{e_i} \cdot \vec{e_1'} \right) \left( \vec{e_j} \cdot \vec{e_1'} \right) +$$
$$+ \frac{1}{2} (\Delta V_\perp)^2 \left[ \left( \vec{e_i} \cdot \vec{e_2'} \right) \left( \vec{e_j} \cdot \vec{e_2'} \right) + \left( \vec{e_i} \cdot \vec{e_3'} \right) \left( \vec{e_j} \cdot \vec{e_3'} \right) \right], \tag{3}$$

where $(\vec{e_1}, \vec{e_2}, \vec{e_3})$ – are the unit vectors of the axes of the laboratory coordinate system; $(\vec{e_1'}, \vec{e_2'}, \vec{e_3'})$, the unit vectors of the axes of the coordinate system connected with the mass center of the approaching stars such that vector $\vec{e_1'}$ is directed along the vector of the relative velocity of approaching stars (see Fig. L.1 in the monograph of Binney and Tremaine, 2008). The variations of the longitudinal and transversal velocity components transformed to the laboratory coordinate system and appearing in formulas (2), (3) for the diffusion coefficient are equal to (ibid., Fig. L.7)

$$\Delta V_\parallel = \frac{2Gm_f p_{90}}{V_0 (p^2 + p_{90}^2)}, \quad \Delta V_\perp = \frac{2Gm_f p}{V_0 (p^2 + p_{90}^2)}, \tag{4}$$

where $m_f, V_0, p$ are the mass of the field star, magnitude of the relative velocity of approaching stars, and impact parameter, respectively.

The variation in the velocity component, $\Delta V_\parallel$, is computed for unit time interval, i.e., it can be treated as an acceleration of the test star due to (stochastic) irregular forces. It therefore seems absolutely logical to multiply $\Delta V_\parallel$ before the integration by $\lambda(p)$, reduction factor of the force. From the other side, $(\Delta V_\parallel)^2$ and $(\Delta V_\perp)^2$ can be treated as the changes of the kinetic energy per unit time and, therefore, these changes are proportional to the *power of force*, and these terms should also be multiplied by $\lambda(p)$, when we further integrate formulas (2) and (3) over impact parameters.

As usual, we integrate the above formulas over impact parameters with the weight $dN(p) = 2\pi \nu V_0 p \, dp$ equal to the number of test-star encounters with field stars with relative velocity $V_0$ and impact parameters in the $(p, p + dp)$ interval over unit time interval. We focus only on the integration over impact parameters, because the subsequent integration over the velocity distribution of field stars yields Rosenbluth potentials (Rosenbluth et al., 1957) like in classical stellar dynamics studies.

*Agekyan's λ-factor* cannot be expressed in terms of elementary functions and therefore we can use our derived piecewise analytical approximations. First, for the sake of convenience (as we show below) we consider *Agekyan's λ-factor* to be a function of $n = N/N_0$, where $N_0 \approx 0.712$ is the average number of stars inside the sphere of radius equal to the average interparticle distance $\bar{d}$. We thus naturally introduce $\bar{d}$ as a scale parameter of the stellar field. We established with simple computations that *Agekyan's λ-factor* can, up to about 2-3%, be approximated by the following simple analytical formulas

$$\lambda(n) \approx \begin{cases} a \cdot exp\left[-b \cdot n^{1/2}\right], & n \leq 1 \\ c \cdot n^{-2/3}, & n > 1 \end{cases} \quad (5)$$

where $a \approx 1.042 \pm 0.001$, $b \approx 1.583 \pm 0.002$, and $c \approx 0.2347 \pm 0.001$ (at a significance level of 95%). This accuracy is quite sufficient for our estimates of integrals.

The current values of the impact parameter and average number of stars in the sphere of the corresponding radius are connected by the following evident relation

$$p = \bar{d} \cdot (N/N_0)^{1/3} = \bar{d} \cdot n^{1/3} \quad (6)$$

The next stage in the computation of the diffusion coefficients consists in integrating velocity changes $\Delta V_\parallel, (\Delta V_\parallel)^2, (\Delta V_\perp)^2$ over impact parameters:

$$DV_\parallel = \pi \nu V_0 \cdot \int_0^\infty \Delta V_\parallel \cdot \lambda(p) \cdot d(p^2) \quad (7)$$
$$DV_\parallel^2 = \pi \nu V_0 \cdot \int_0^\infty (\Delta V_\parallel)^2 \cdot \lambda(p) \cdot d(p^2) \quad (8)$$
$$DV_\perp^2 = \pi \nu V_0 \cdot \int_0^\infty (\Delta V_\perp)^2 \cdot \lambda(p) \cdot d(p^2) \quad (9)$$

It is evident that $DV_\parallel^2 \ll DV_\perp^2$, because integral (8) converges in classical computations of the diffusion coefficients, and the convergence is even more evident in our case where the integrand is multiplied by a rapidly decreasing function of impact parameter. That is why we do not consider diffusion coefficient (8) below.

We now use formula (6) and pass from integration over impact parameter to integration over the relative number of stars $n$ by transforming formula (7) to the form

$$DV_\parallel = \frac{2\pi G^2 m_f (m+m_f)\nu}{V_0^2} \cdot \int_0^\infty \frac{\lambda(p) \, d(p^2)}{(p^2+p_{90}^2)} = \frac{2\pi G^2 m_f (m+m_f)\nu}{V_0^2} \cdot K^2 \cdot \int_0^\infty \frac{\lambda(t) \, dt}{1+K^2 t} =$$
$$= \frac{2\pi G^2 m_f (m+m_f)\nu}{V_0^2} \cdot I_1(K) \quad (10)$$

where $K = \bar{d}/p_{90}$ is the ratio of two scale lengths of the stellar field and the new integration variable is $t = (N/N_0)^{2/3}$. We similarly derive the following formula for the quadratic diffusion coefficient

$$DV_\perp^2 = \frac{4\pi G^2 m_f^2 \nu}{V_0} \int_0^\infty \frac{\lambda(p) \, p^2 \, dp}{(p^2+p_{90}^2)^2} = \frac{4\pi G^2 m_f^2 \nu}{V_0} \cdot K^4 \cdot \int_0^\infty \frac{\lambda(t) \, t \, dt}{(1+K^2 t)^2} =$$
$$= \frac{4\pi G^2 m_f^2 \nu}{V_0} \cdot I_2(K), \quad (11)$$

where the dimensionless functions

$$I_1(K) = K^2 \cdot \int_0^\infty \frac{\lambda(t) \, dt}{1+K^2 t}, \quad I_2(K) = K^4 \cdot \int_0^\infty \frac{\lambda(t) \, t \, dt}{(1+K^2 t)^2} - \quad (12)$$

appearing in formulas (10) and (11) for the first- and second-order diffusion coefficients, respectively, depend only on scale factor ratio $K$.

We computed functions (12) by numerically integrating the corresponding integrands for a wide range of scale factor ratios $1 < K < 10^5$. Note that the upper boundary of parameter $K$ corresponds to rather low star number density on the order of 0.1 pc$^{-3}$, which resembles the conditions in the solar neighborhood. Figure 2 shows the behavior of integral $I_1$ with increasing upper integration limit for the scale factor ratio of $K = \bar{d}/p_{90} = 1000$. It is evident from the figure that the function levels off already at small t values demonstrating the total absence of logarithmic divergence. We now recall that $t = \left(d_{max}/p_{90}\right)^2$ to see that the integral saturates and the test star becomes practically "shielded" from distant encounters at distances as small as 2–3 average interparticle distances.

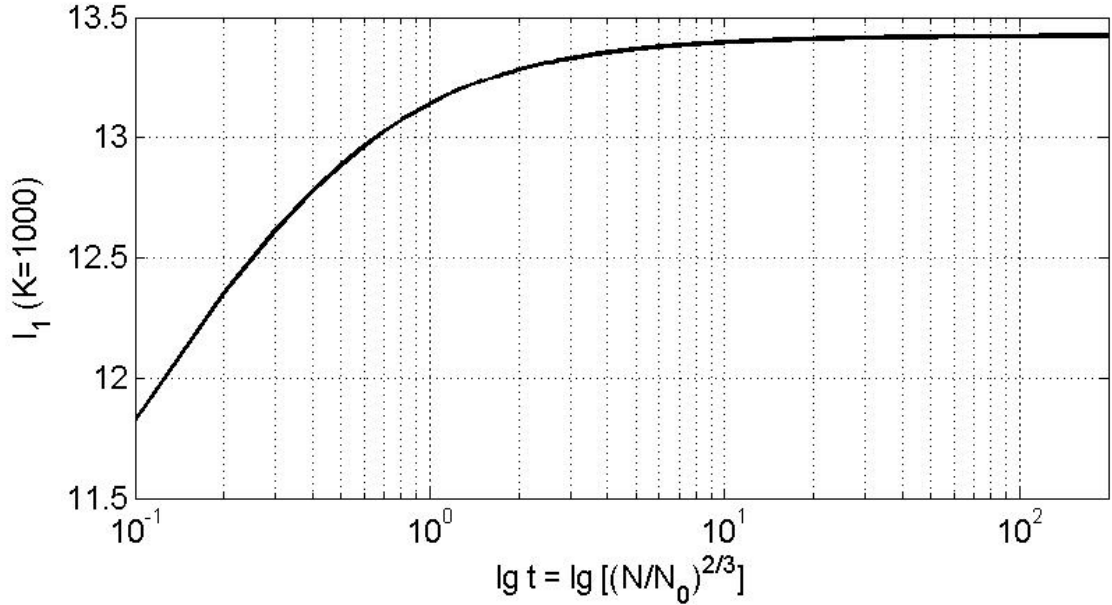

Fig. 2. Behavior of integral $I_1(K = 1000)$ in formula (10) as a function of the upper integration limit, $t_{max}$.

Figure 3 shows the behavior of integral $I_2$ as a function of the upper integration limit for the same scale ratio $K = \bar{d}/p_{90} = 1000$. It is evident from the figure that this integral converges even faster, and the test star becomes actually "shielded" from distant encounters at distances as small as about 1–2 average interparticle distances. This is no surprise given the very rapid decrease of *Agekyan's λ-factor* with impact parameter of the encounters.

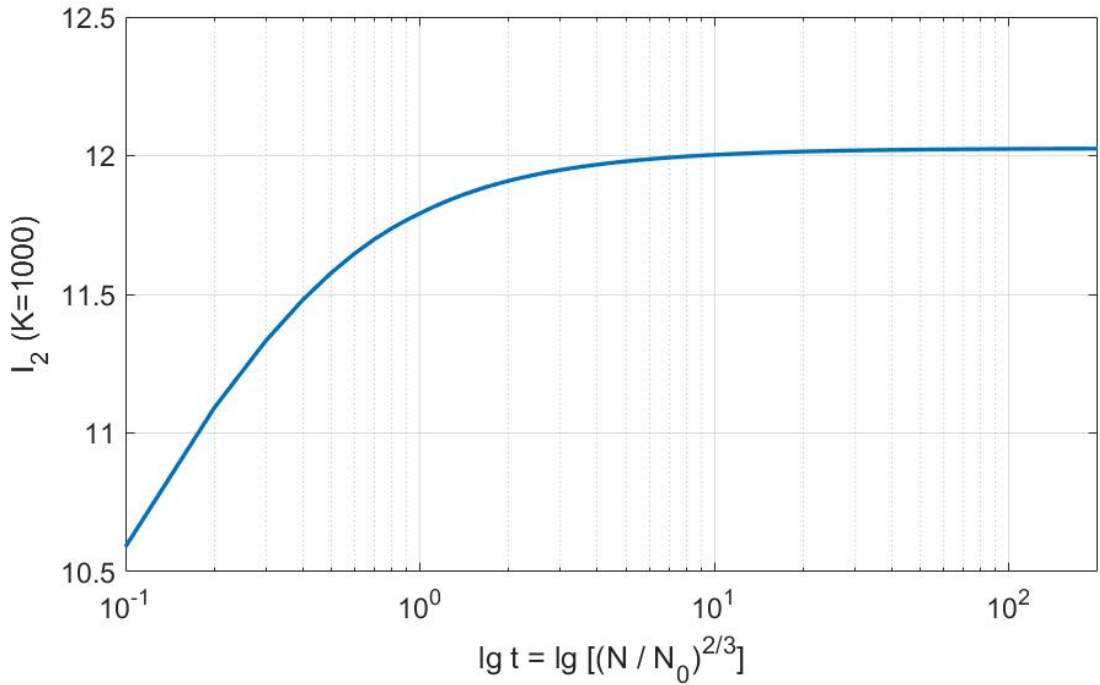

Fig. 2. Behavior of integral $I_2(K = 1000)$ in formula (11) as a function of increasing upper integration limit, $t_{max}$.

Figures 4 and 5 show the behavior of integrals $I_1$ and $I_2$ as functions of parameter $K$.

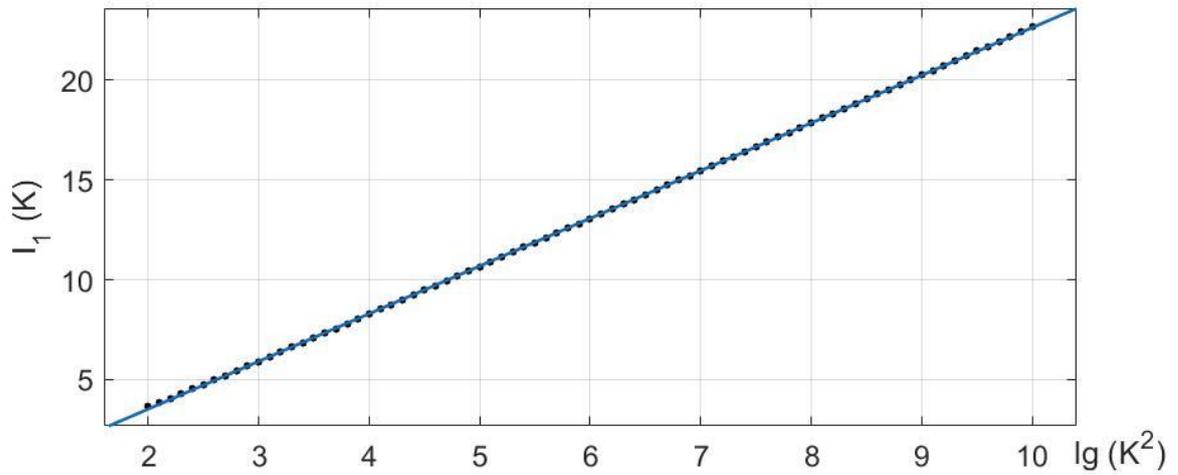

Fig. 4. Dependence of integral $I_1$ in formula (10) on parameter $K$ in the interval $10 < K < 10^5$. The dots show the results of computations with a constant step on $\lg(K^2)$; the solid line shows very good linear approximation in decimal logarithm, $\lg(K^2)$, and the dashed lines, the 95% confidence limits.

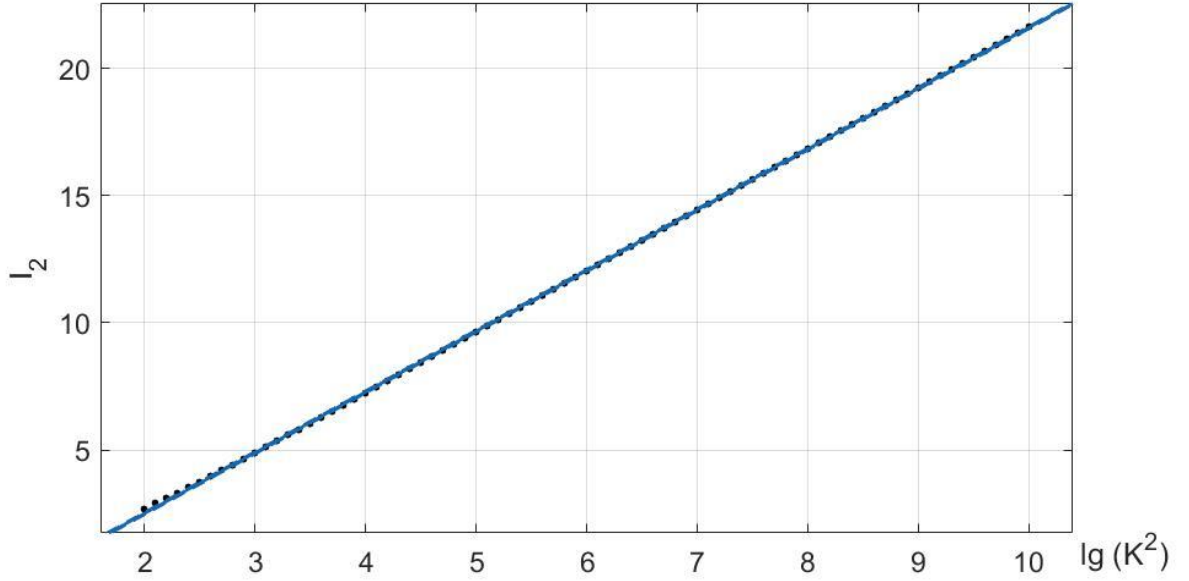

Fig. 4. Dependence of integral $I_2$ in formula (12) on parameter $K$ in the interval $10 < K < 10^5$. Designations are the same as in Fig. 4. It is evident that the computed values and their linear approximations practically coincide and the deviations do not exceed the sizes of the symbols.

The parameters of linear dependences $I_1$ and $I_2$ on $lg\ (K^2)$ for sufficiently large parameter values, $K > 10,$ can be easily derived from the results of our computations, which we show in the plots in Figs 4 and 5:

$$I_1(K) \approx (2.39 \pm 0.01) \cdot lg(K^2) - (1.26 \pm 0.03) \approx 2.07 \cdot ln(K/1.8) \qquad (13)$$

$$I_2(K) \approx (2.39 \pm 0.04) \cdot lg(K^2) - (2.26 \pm 0.02) \approx 2.07 \cdot ln(K/3.0), \qquad (14)$$

where $ln$ symbol is used for natural logarithm. We now substitute equations (13) and (14) into equations (10) and (11), respectively, to obtain the final formulas for the diffusion coefficients with the allowance for gravitational "shielding" of distant two-particle encounters:

$$DV_\parallel \approx \frac{4.15\ \pi G^2 m_f(m+m_f)v}{V_0^2} \cdot ln\left(\bar{d}/1.8 \cdot p_{90}\right), \qquad (15)$$

$$DV_\perp^2 \approx \frac{8.30\ \pi G^2 m_f^2 v}{V_0} \cdot ln\left(\bar{d}/3.0 \cdot p_{90}\right) \qquad (16)$$

The coefficients in (13-14) are close to 2; we are sure that for more accurate approximation for Agekyan's factor these values will be equal exactly to 2.

## 3. Discussion of results

Let us now compare our computed diffusion coefficients with the results of the classical computations with the effect of distant encounters intuitively cut off at the average interparticle distance (see, e.g., Williamson and Chandrasekhar, 1941):

$$DV_\parallel \approx \frac{4\pi G^2 m_f(m+m_f)\nu}{V_0^2} \cdot ln\left(\bar{d}/p_{90}\right), \qquad (17)$$

$$DV_\perp^2 \approx \frac{8\pi G^2 m_f^2 \nu}{V_0} \cdot ln\left(\bar{d}/e \cdot p_{90}\right) \qquad (18)$$

First, given that we are concerned only with estimating the effect of binary stellar encounters, it is safe to say that the use of formulas (15) and (17), (16) and (18) for practical computations of diffusion and the time scales does not bring any large discrepancies. Second, from Fig. 2 and 3 it becomes absolutely clear that the effective maximum impact parameter in terms of the two-particle encounter concept should indeed be not significantly larger than 1–2 average interparticle distances, and all the more distant encounters contribute mostly to the regular force component. We thus actually corroborated the point of view of the researchers who consider it necessary to restrict the effect of irregular forces calculated in the frame of binary encounters by the average interparticle distance. Strictly speaking, encounters with $p \geq (5-10)\bar{d}$ do not contribute to the irregular force at all.

Third, and most important, both our formulas and those proposed in classical works contain the logarithmic factor. However, in our approach it has a fundamentally different physical meaning. We show that the allowance for the multiplicity of encounters allows avoiding the divergence of integrals at the upper limit. In our case the logarithmic factor appears naturally and is due to the fact that any stellar medium is characterized by two totally independent scale lengths: the average interparticle distance, $\bar{d} \approx 0.554 \cdot \nu^{-1/3}$, which is related only to the concentration of stars, and the parameter of close encounter, $p_{90} = \frac{G(m+m_f)}{V_0^2}$, which reflects the dynamics of the stellar medium (it is determined by the masses and characteristic velocities of stars). It is important that these parameters become directly related only under the conditions of virial equilibrium.

Nearly the same result was obtained earlier by Kandrup (1981) who analyzed kinetic processes in a locally homogeneous stellar media (e.g. homogeneous over distance comparable to the local mean interparticle spacing). Using the distribution of random forces analogous to Holtsmark, he was able to derive rigorous expressions for the diffusion coefficients (formulas 139–140 from his paper) completely similar to classic expressions for uniform infinite stellar field. He also emphasized that the Coulomb logarithm in these expressions does reflect the ratio of two characteristic scales rather than logarithmic divergence. Finally, we must note that two different methods used by us and by Kandrup (1981) to calculate the contribution of random forces to the diffusion in the velocity space, lead to the same conclusion about very effective shielding of distant encounters, which result in the convergence of diffusion coefficients, in contrast to classical artificial cut-off of distant encounters.

*Agekyan's λ-factor* was derived based on the Holtsmark distribution for uniform stellar medium. In real stellar systems the size of spatial irregularities (density fluctuations) is significantly greater than the average interparticle distance, and therefore the estimates of the range of *irregular forces* produced by stars obtained in this paper are, in our opinion, quite applicable to nonuniform systems as well, as was also shown by Kandrup (1981). It is evident that the influence of spatial irregularities may show up as collective effects in the gravitating medium, including the effects due to the fractal structure of the medium (Chumak and Rastorguev, 2015). Vlad (1994), Chavanis (2009), and Chumak and Rastorguev (2016) showed that the distribution of random force in a fractal medium can be described by a full analog of the Holtsmark distribution, where the average star number density is replaced by the conditional density computed based on the observed fractal dimension of the medium. We believe that the passage to the limit in the computation of kinetic coefficients in the case of fractal medium also allows us to ignore the effect of irregularities located beyond several *intercluster* distances.

## Acknowledgments

O.V. Chumak acknowledges the support from the Russian Foundation for Basic Research (grant no. № 14-02-00472) and A.S. Rastorguev acknowledges partial support from the Russian Science Foundation (grant no. 14-22-00041). Authors are also grateful to Dr. P. Chavanis, Dr. V.Yu. Terebizh and to anonymous referee for valuable notes.